\documentclass[useAMS,usenatbib,twocolumn]{mn2e}
\usepackage{hyperref}
\usepackage{graphicx}
\usepackage{multirow}
\usepackage{amsmath}
\usepackage{times}
\usepackage{makecell}
\usepackage{caption}

\title{Finding the imprints of stellar encounters in long period comets}
\author[Fabo\ Feng \& C. A. L.\ Bailer-Jones]
{Fabo Feng \& C. A. L.\ Bailer-Jones\\
  Max Planck Institute for Astronomy, K\"onigstuhl 17, 69117 Heidelberg}
\date{\today}

\begin{document}


\maketitle

\begin{abstract}
  The solar system's Oort cloud can be perturbed by the Galactic tide and by individual passing stars. These perturbations can inject Oort cloud objects into the inner parts of the solar system, where they may be observed as the long-period comets (periods longer than 200 years). Using dynamical simulations of the Oort cloud under the perturbing effects of the tide and 61 known stellar encounters, we investigate the link between long-period comets and encounters.  We find that past encounters were responsible for injecting at least 5\% of the currently known long-period comets. This is a lower limit due to the incompleteness of known encounters. Although the Galactic tide seems to play the dominant role in producing the observed long-period comets, the non-uniform longitude distribution of the cometary perihelia suggests the existence of strong -- but as yet unidentified -- stellar encounters or other impulses. The strongest individual future and past encounters are probably HIP 89825 (Gliese 710) and HIP 14473, which contribute at most 8\% and 6\% to the total flux of long-period comets, respectively. Our results show that the strength of an encounter can be approximated well by a simple proxy, which will be convenient for quickly identifying significant encounters in large data sets. Our analysis also indicates a smaller population of the Oort cloud than is usually assumed, which would bring the mass of the solar nebula into line with planet formation theories.
\end{abstract} 

\section{Introduction}

Planetary craters reveal an extensive history of bombardments caused by asteroids and comets. Large impacts on the Earth may have triggered global catastrophic events such as the K-T mass extinction \citep{alvarez80,schulte10}. To understand such events and to predict the risk of future impacts it is necessary to study the past and current impact rate in the solar system.  One approach to this is to date impact craters \citep{ivanov02,bottke05}, although many terrestrial craters have large age uncertainties and very few ages exist for extraterrestrial craters \citep{wilhelms87,culler00,jourdan12}. Another approach is to dynamically simulate the motions of asteroids and comets and to calculate their impact rates. For example, the late heavy bombardment can be explained by the migration of giant planets in the Nice model \citep{gomes05,tsiganis05}.

In addition to the analysis of minor bodies in the inner solar system, many previous studies have examined the evolution of the outer solar system and the planetary impact rate associated with it \citep{everhart67,weissman07,kaib09}. This may be modulated by the solar orbit in the Galaxy, as this changes the stellar density in the solar neighbourhood \citep{gardner11,feng13,feng14}. Specifically, the outer part of the solar system is occupied by the Oort cloud (OC), which was proposed by \cite{oort50} to explain the large semi-major axes ($>20,000$\,au) of long-period comets (those with periods larger than 200\,yr). The angular momenta of comets in the OC are changed by the gravitational perturbations from the Galactic tide and stellar encounters. When the angular momentum or perihelion of an OC comet becomes small enough, it would be observed and classified as a long-period comet (LPC). 
By integrating the orbit of an LPC back in time we may be able to identify the stellar encounter responsible.
\citep{dybczynski02,rickman12, feng14}. In general, dynamical simulations of the formation and evolution of the OC are crucial for improving our understanding of the origin of Sedna-like bodies \citep{brown04, morbidelli04,kenyon04,kaib11,trujillo14}, the birth environment of the Sun \citep{levison10,brasser12}, and the cometary impact rate on the Earth \citep{feng14}.

However, few studies have systematically investigated the influence of observed stellar encounters on the OC and thus on the flux of LPCs. Most studies have only estimated the influence of a few close encounters
or have used empirical models, rather than through dynamical simulations. 
In this work we try to better quantify the influence of encounters by using dynamical simulations of the OC to model the cumulative effect of many close encounters. We collect known close and strong encounters from \cite{bailer-jones15,dybczynski15b,mamajek15}. We then simulate the OC comets by taking into account the gravitational perturbations from these encounters and the Galactic tide. To investigate the role of each perturber, we conduct simulations of the OC under perturbations from the tide only, from encounters only, and from both (which we call ``combined simulations''; see \citealt{feng14} for details). We use subscripts $t$, $e$, $c$ to refer to these types of simulations respectively. 
Using these results we measure the effect of each encounter using various metrics. By comparing the results of the three types of simulations we investigate the nature of non-linear interactions between the tide and encounters, the so-called``synergy effect'' \citep{heisler87,rickman08}. Finally, we compare the angular distributions of the perihelia of LPCs and encounters to assess the influence of the solar motion on the current LPC flux. 

\section{Data}\label{sec:data}

We collected the data of stellar encounters from three sources: \cite{bailer-jones15} (hereafter BJ15), \cite{dybczynski15b} and \cite{mamajek15}. Following \cite{bailer-jones15}, we use the term ``object'' to refer to each encountering star in our catalogue. A specific star may appear more than once but with different data, thus leading to a different object. As described in BJ15, we use GCS (G), RAVE (R), Pulkovo (P) to denote the catalog of objects with radial velocity obtained by cross-matching the revised Hipparcos catalog \citep{leeuwen07} with the Geneva-Copenhagen survey data by \citep{casagrande11}, with the Pulkovo catalogue \citep{gontcharov06}, and with Rave-DR4 \citep{kordopatis13}, respectively. We also use data from the XHIP (X) catalog \citep{anderson12} and the XHIP catalog adapted by \citep{dybczynski15b} (D). We also include WISE J072003.20-084651.2 (W0720) with data from \citep{scholz14,mamajek15} (M). In summary, the catalogs of G, R, P and X are from BJ15, and catalog D and M are from their corresponding references.

We first select all objects with mean perihelia $d_{\rm ph}<2$\,pc from the catalog provided by BJ15. This gives 65 objects. However, more distant encounters can also have significant perturbing effect. We therefore select additional encounters using a proxy, $g$, for the size of the perturbation, defined as
\begin{equation}
  g = \frac{M_{\rm enc}}{d_{\rm ph}^2 v_{\rm ph}}=\frac{\gamma}{d_{\rm ph}}\,
  \label{eqn:proxy}
\end{equation}
where $M_{\rm enc}$ is the mass of the encountering star in solar masses, $d_{\rm ph}$ is its perihelion distance in pc, and
$v_{\rm ph}$ is its velocity at perihelion in the heliocentric rest frame \citep{rickman76} in km$/$s. $\gamma$ (which is defined by the above equation) is linearly proportional to the impulse of the Sun gained from the encounter, and was used as a proxy for encounter strength in our previous work \citep{feng14} (hereafter FBJ14). We will see in section \ref{sec:individual} that $g$ is actually a better proxy, which is why we use it here.

The values of $g$ for the 65 selected objects ranges from $10^{-4}$ to 3. Of these, 34 objects have $g<0.01(M_\odot{\rm pc}^{-2}{\rm km}^{-1} {\rm s})$, and these can only inject about one in a million LPCs from the OC comets (see section \ref{sec:individual} and Table \ref{tab:encounter}). 
To select encounters with at least this perturbation strength but which have perihelia beyond 2\,pc, we additionally select all encounters with $g>0.01(M_\odot{\rm pc}^{-2}{\rm km}^{-1} {\rm s})$, yielding another 25 objects from BJ15's catalog. We then also select the 12 new encountering objects with perihelia less than 2\,pc found by \cite{dybczynski15b}, as well as the close encounter W0720 from \cite{mamajek15}. The masses of these objects are either from the literature or estimated according to their stellar types based on Table 1 of \cite{rickman08}. Since many sources may be unresolved binaries, the masses may have uncertainties as large as 200\%, especially for the estimated masses. We are left with a sample of 103 objects, of which 90 are from BJ15 and 13 are from other catalogs. We will do full integrations for all selected sources. 

For each object, we resample the observed data according to the observational errors (as described in BJ15) to create 100 different initial conditions. We integrate the orbits of all samples, and select the trajectory which gives a perihelion closest to the mean perihelion given in the catalogs \footnote{The mean and median of the distributions of perihelion time, distance and velocity are very similar: see BJ15 and its supplementary information on \url{http://www.mpia.de/homes/calj/stellar_encounters/stellar_encounters_1.html}}. This trajectory is more representative of the mean perihelion than using the nominal observed data, due to the nonlinear propagation of errors (see BJ15). The time ($t_{\rm ph}$), velocity ($v_{\rm ph}$), and distance ($d_{\rm ph}$) of each object at its perihelion are calculated for the selected trajectory. The resulting data are shown in Table \ref{tab:encounter}.

In what follows we study the individual and collective influence of the 103 encountering objects on the OC. For this collective assessment we select a unique object for each star to generate a sequence of encounters such that each star occurs only once. We select objects from the catalogs in the following decreasing order of priority (set by our assessment of their reliability): G, P, X, M, D, and R. We call this set of object the {\em basic sequence}. Some objects in this sequence have dubious radial velocities, proper motions, or parallaxes. By excluding these objects according to the comments in BJ15 and \cite{dybczynski15b}, we define a subset of the basic sequence which we call the {\em reliable sequence}. Encounters are labelled ``b'' or ``r'' in Table \ref{tab:encounter} according to which sequence they belong to (all ``r'' are also in the basic sequence), while those labelled``n'' are not in either sequence.

\begin{table*}
    \begin{center}
\caption{Data on the 45 most perturbing encounters according to the number of encounter-induced LPCs, $N_e$ (and sorted by this value). This value was
calculated from the encounter-only simulations using one million OC comets. 
The objects are denoted using Hipparcos IDs except for W0720, which is denoted by ``0720''. 
The penultimate column (``seq'') denotes the sequence which an object belongs to: ``r'' denotes objects in the reliable sequence, a subset of the basic sequence with reliable data; ``b'' means objects in the basic sequence but not in the reliable sequence; and ``n'' indicates objects in no sequence. The reference to the stellar masses is given in the last column: ``(1)-estimated'' means the mass was estimated according to the stellar type. The full list of 103 encounter objects is available online. }
\label{tab:encounter}
\hspace{-0.2in}  \begin{tabular}{lcccc cccc cccc}
  \hline
  \hline
  HIPID & cat & $M_{\rm enc}$ & $t_{\rm ph}$&$v_{\rm ph}$ & $d_{\rm ph}$ & $g\times 100$ & $N_e$ & $f_{\rm max}$&$\Delta T_{\rm max}$& $f_0$ & seq & ref\\
        &     & $M_\odot$    & Myr       & km\ s$^{-1}$  & pc         & $M_\odot{\rm pc}^{-2}{\rm km}^{-1} {\rm s} $ & &$\times 10^3$&Myr&$\times 10^4$&&\\
  \hline
85605&X&0.7&0.34&20.98&0.103&312.0&539&555.3&0.80&0.0&b&(1)\\
89825&X&0.6&1.42&13.99&0.267&60.4&187&154.6&1.19&0.0&n&(2)\\
63721&P&1.7&0.14&36.09&0.273&63.2&159&137.5&1.30&0.0&b&(1)\\
89825&P&0.6&1.30&14.76&0.363&30.8&115&82.0&1.56&0.0&r&(2)\\
14473&D&1.1&-3.52&36.16&0.282&38.4&67&58.4&1.79&362.1&r&(1)\\
14576&X&6.0&-8.48&3.28&4.059&11.2&38&21.5&1.43&38.0&n&(3)\\
103738&P&2.5&-3.71&18.09&0.825&20.3&31&14.9&9.33&61.0&r&(4)\\
14576&P&6.0&-4.30&6.48&3.145&9.4&31&18.0&1.34&114.4&r&(3)\\
85661&P&1.7&1.92&46.39&0.594&10.4&22&9.7&5.10&0.0&r&(1)\\
84263&D&1.0&-6.29&11.92&1.201&6.0&14&7.6&0.80&36.5&r&(1)\\
27288&X&3.2&-0.86&24.76&1.300&7.6&12&7.8&5.14&45.4&n&(1)\\
27288&P&3.2&-0.86&24.50&1.300&7.7&12&7.8&5.07&45.4&r&(1)\\
26744&D&1.3&12.64&4.83&1.714&9.2&9&2.2&9.31&0.0&n&(1)\\
25001&P&3.1&4.92&17.00&2.185&3.8&8&6.4&13.86&0.0&r&(1)\\
71683&G&1.1&0.03&34.14&0.902&4.0&7&5.1&0.84&0.0&r&(5)\\
71683&P&1.1&0.03&34.29&0.910&3.9&7&5.1&0.84&0.0&n&(5)\\
71683&X&1.1&0.03&31.84&1.001&3.4&7&5.1&0.84&0.0&n&(5)\\
87052&D&3.2&5.75&35.68&1.812&2.7&7&5.1&0.88&0.0&r&(1)\\
71681&R&0.9&0.03&29.67&0.912&3.6&6&5.1&0.84&0.0&n&(5)\\
25001&D&3.1&3.97&22.83&2.069&3.2&6&5.7&14.47&0.0&n&(1)\\
30344&P&0.8&-1.94&14.71&1.394&2.7&5&3.7&17.66&7.5&n&(1)\\
71681&X&0.9&0.03&30.32&1.062&2.6&4&3.7&19.55&0.0&r&(5)\\
30344&G&0.8&-1.98&14.25&1.415&2.7&4&3.7&17.67&6.5&r&(1)\\
26624&X&2.1&-1.84&22.89&1.939&2.4&4&3.7&17.78&8.8&n&(1)\\
26624&P&2.1&-1.88&22.06&1.984&2.4&4&3.7&17.74&7.9&r&(1)\\
25001&X&3.1&3.64&23.47&2.226&2.7&4&3.5&3.45&0.0&n&(1)\\
42525&X&1.3&-0.26&59.99&0.955&2.4&3&2.1&6.26&20.3&r&(1)\\
94512&X&3.2&3.77&30.21&1.829&3.2&3&2.1&9.09&0.0&n&(1)\\
93506&X&5.3&-1.01&25.90&3.323&1.8&3&4.1&0.45&35.6&n&(6)\\
93506&P&5.3&-1.02&26.03&3.329&1.8&3&4.1&0.43&34.7&r&(6)\\
23415&D&3.2&4.22&30.90&1.650&3.8&3&2.1&0.16&0.0&r&(1)\\
91012&R&1.3&0.26&349.23&0.486&1.6&2&4.2&5.90&0.0&b&(1)\\
30344&X&0.8&-1.55&18.86&1.160&3.1&2&2.1&17.31&0.0&n&(1)\\
38965&X&1.3&-1.23&54.83&1.330&1.3&2&2.1&0.02&10.0&r&(1)\\
25240&P&1.5&-1.06&54.60&1.623&1.1&2&2.1&0.44&17.2&n&(1)\\
25240&G&1.5&-0.99&53.82&1.665&1.0&2&2.1&0.50&18.3&r&(1)\\
94512&P&3.2&3.51&29.59&2.241&2.2&2&2.1&14.40&0.0&r&(1)\\
32349&P&3.0&0.04&17.54&2.505&2.7&2&2.1&0.48&0.0&r&(7)\\
32349&X&3.0&0.05&17.55&2.498&2.7&2&2.1&0.48&0.0&n&(7)\\
26744&X&1.3&12.98&5.19&2.630&3.6&2&1.1&-2.90&0.0&n&(1)\\
26744&P&1.3&16.13&3.94&2.990&3.7&2&1.1&-6.05&0.0&n&(1)\\
26744&G&1.3&13.16&4.74&2.969&3.1&2&1.1&-3.08&0.0&r&(1)\\
18907&X&2.2&5.00&6.93&5.107&1.2&2&3.9&1.14&0.0&n&(8)\\
18907&P&2.2&4.96&7.05&5.698&1.0&2&3.9&1.18&0.0&r&(8)\\
W0720&M&0.1&-0.07&83.58&0.252&2.8&1&5.2&0.44&52.3&r&(9)\\
\hline
\end{tabular}
\\[3.0pt]
(1)-estimated, (2)-\cite{sanchez99}, (3)-\cite{rhee07}, (4)-\cite{bailer-jones15}, (5)-\cite{thevenin02},\\ (6)-\cite{derosa12}, (7)-\cite{liebert05}, (8)-\cite{malagnini90}, (9)-\cite{mamajek15}
\end{center}
\end{table*}

\section{Method}\label{sec:method}

\subsection{Simulation scheme}\label{sec:simulation}

Following the simulation scheme in FBJ14, we conduct three sets of simulations to study the influence of stellar encounters on the OC: tide-only, encounter-only, and combined encounter-tide simulations using the {\small AMUSE} software \citep{portegies09,pelupessy12}. We use the Galaxy model and initial conditions of the Sun as in BJ15. The initial conditions of the OC comets are from FBJ14, which were generated from a model adapted from \cite{dones04b}'s results, and which we call the DLDW model. In this we used an inner boundary for the smallest perihelia of comets in order to avoid the influence of planets on the OC. However, this results in a non-equilibrium initial distribution. To overcome this problem, we use tide and encounter-like perturbations (see FBJ14) over 100\,Myr to move the OC closer to the kind of steady state it would have around the present time. We then take the coordinates of 10 million comets which remain in the OC at the end of this as the initial conditions for the following simulations. We will integrate the orbit of each individual OC comet under the influence of the gravitational perturbations. 

In the following we apply the tide-only, encounter-only, and combined simulations to the time span covering each individual stellar encounter as well as the two sequences of encounters. For the study of individual encounters we simulate one million OC particles over a time span of 10\,Myr or 20\,Myr 
from now, either into the future or the past according to when the perihelion occurs. Specifically, if $|t_{\rm ph}|>10$\,Myr we adopt a 20\,Myr time span, otherwise we use a 10\,Myr. In the case of encounter sequences, we simulate 10 million OC comets over a time span of 10\,Myr. Encounters with perihelion times out of this window are naturally excluded.  

\subsection{Analysis of LPC flux}\label{sec:analysis}

Having integrated the comet orbits in the above simulations, we then calculate the resulting LPC flux. To do this we use two definitions of an LPC.

First, an OC comet is counted as an LPC when its perihelion is less than 15\,au. Within this so-called ``loss cone'' (LC), planetary perturbations can strongly influence the orbit of the comet and either capture or eject it \citep{wiegert99}. Although {\em observed} LPCs have perihelia less than 5\,au (the ``observable zone''), we found in earlier work (FBJ14) that the flux of comets injected into the LC is approximately linearly proportional to that injected into the observable zone. The injection time of an LPC is when its pericentric distance is equal to $15$\,au, $r=q_{\rm lc}$. This injection time is calculated by extrapolating the trajectory of the comet from the place where it gains the impulse to its perihelion, assuming a Keplerian motion. 

In addition to LPCs in the LC (or LC LPCs), there is another type of LPC, ``dynamically new LPCs''. These are comets which have enough energy to penetrate the Jupiter--Saturn barrier and enter the observable zone within one revolution \citep{heisler86}. Following \cite{dybczynski15a}, we count an LPC as a new LPC when its perihelion decreases from a value larger than 15\,au to a value less than 5\,au within one revolution. Since planetary perturbations probably cannot change a new LPC's perihelion significantly over one revolution, our neglect of planetary perturbations in the simulations does not significantly diminish the accuracy of the computed LPC flux.

\begin{figure}
  \centering
  \includegraphics[scale=0.7]{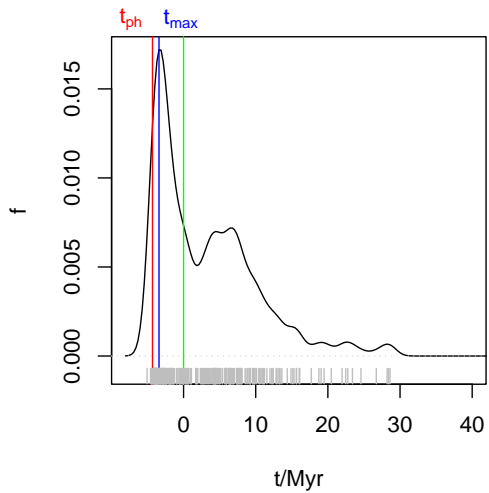}
  \caption{The distribution of the flux ratio, $f$, over time for HIP 14576 (Algol) based on backward simulations of 1 million OC comets over 10\,Myr. 
This distribution is obtained from a kernel density estimate over the injection times of the individual LPCs (shown in grey lines at the bottom) using a Gaussian kernel with a scale length of 1\,Myr. 
The green and red lines indicate $t=0$\,Myr and $t=t_{\rm ph}$ (encounter perihelion) respectively. The blue line marks the time when $f$ reaches its maximum. }
  \label{fig:density}
\end{figure}

With these definition of LPCs we calculate the injection time of LPCs generated in simulations of the OC. 
We use kernel density estimation to convert these discrete times into a continuous LPC flux as a function of time, $F_e(t)$. From this we can estimate the maximum and the current LPC flux induced by individual encounters. To compare this with the effect of the Galactic tide, we estimate the strength of an encounter using the ratio of encounter-induced to tide-induced LPC flux. 

Since the tidal force imposed on the OC from the Galaxy hardly changes over a few million years (FBJ15), the LPC flux caused by the tide, $F_t(t)$, can be considered as constant, i.e.\ $F_t(t)=F_t$ over our integration time span of 10\,Myr.
To quantify the influence of encounters on the Oort cloud, we define the flux ratio as
\begin{equation}
  f(t) = F_e(t)/F_t\ ,
\label{eqn:f}
\end{equation} 
where $F_t$ is the average flux over the integration time span. The encounter-induced flux, $F_e(t)$, is derived from the encounter times using a Gaussian kernel with a scale length of 1\,Myr. Since the tide dominates the production of current LPCs \citep{matese11,dybczynski15b}, the flux ratio is also a good approximation of the contribution of encounters to the total LPC flux. 

To characterize the perturbations on the OC from encounters, we further define the maximum flux ratio as
\begin{equation}
  f_{\rm max} = F_e(t_{\rm max})/F_t\ ,
  \label{eqn:fmax}
\end{equation}
where $F_e(t_{\rm max})$ is the maximum flux of the encounter-induced LPCs, and $t_{\rm max}$ is the corresponding time. If the encounter-induced LPC flux is small, $f_{\rm max}$ may be sensitive to the scale length used to smooth the temporal distribution of LPCs. We should exercise caution in interpreting $f_{\rm max}$ when the number of encounter-injected LPCs, $N_e$, is less than 10. From this we also define the time delay, $\Delta T_{\rm max}$, between the time when encounter-induced flux reaches its maximum and the perihelion time, $t_{\rm ph}$. 

To estimate the influence of encounters on the current LPC flux, we further define the current flux ratio as
\begin{equation}
  f_0 =  F_e(t=0)/F_t\ .
  \label{eqn:f0}
\end{equation}
Since an encounter can only significantly influence the OC at its perihelion -- very slow encounters being rare exceptions -- future encounters generally do not influence the current LPC flux. Thus $f_0$ is forced to be zero for future encounters. The calculation of $f_{\rm max}$, $\Delta T_{\rm max}$ and $f_0$ is illustrated by the example of Algol in Figure \ref{fig:density}.

\section{Results}\label{sec:results}

\subsection{LPCs induced by individual encounters}\label{sec:individual}

We conduct tide-only, encounter-only, and combined simulations over 10\,Myr or 20\,Myr from now for each encounter for one million OC comets generated from the DLDW model. We count the resulting number of LPCs as a function of time, analyze the influence of encounters using the statistics introduced in section \ref{sec:analysis}, and show the results in Table \ref{tab:encounter}. Among the 103 objects in our encounter sample, we only show the quantities of the 44 strongest objects which induce at least 2 LPCs in the simulations. We also show the results for W0720, which is one of the closest stellar encounters found so far \citep{mamajek15}.

In table \ref{tab:encounter} there are 12 objects which induce more than 10 LPCs. Of these, seven induce LPCs which have perihelion times around $t=0$\,Myr. We select all the objects in the reliable sequence and add their current flux ratio, $f_0$, and find that these objects induce about 5\% of the current LPC flux. Although the first three objects have strong influence on the Oort cloud, their data are unreliable (BJ15 and \citealt{dybczynski15b}), so we do not discuss their results further. We find that the encounter which has reliable data and induces the most LPCs is HIP 89825 (GL 710). This object is also confirmed as the strongest near-future encounter by many previous studies \citep{sanchez99,bobylev10b}. However, it can induce an LPC flux which is only 8\% of the tide-induced flux. In contrast, \citep{matese02} has found GL 710 to induce 1.4 times as many LPCs as the tide. This discrepancy is understandable because we calculate the time when comets actually reach their perihelia, $t(r=q_{\rm lc})$, while \citep{matese02} use the time when comets achieve the necessary impulse to (later) reach a small perihelion, $t(q=q_{\rm lc})$. Another way of seeing this is that we calculate the flux by effectively averaging the number of LPCs over the typical orbital period of an LPC (about 2\,Myr) while they effectively average over the encounter time scale (about 0.1\,Myr). 

Since HIP 85605 and 63721 do not have reliable astrometric solutions (BJ15 and \citealt{dybczynski15b}), we instead consider GL 710 and 85661 as the strongest two future encounters. The strongest past encounter is HIP 14473, contributing about 3.6\% to the current LPC flux, and at most 6\% of the total LPC flux. We note, however, that this star has larger proper motion uncertainties \citep{dybczynski15b}. We find that the other two strong past encounters are HIP 103738 (gamma Microscopii) and HIP 14576 (Algol), both of which are massive and relatively slow. The radial velocity of Algol is very close to zero with a standard error similar to its magnitude, leading to a large uncertainty in the perihelion time. Our results are largely consistent with the findings of \cite{fouchard11}, namely that massive stars can have strong effects on the Oort cloud without having to make a very close encounter.

\begin{figure}
  \centering
  \includegraphics[scale=0.6]{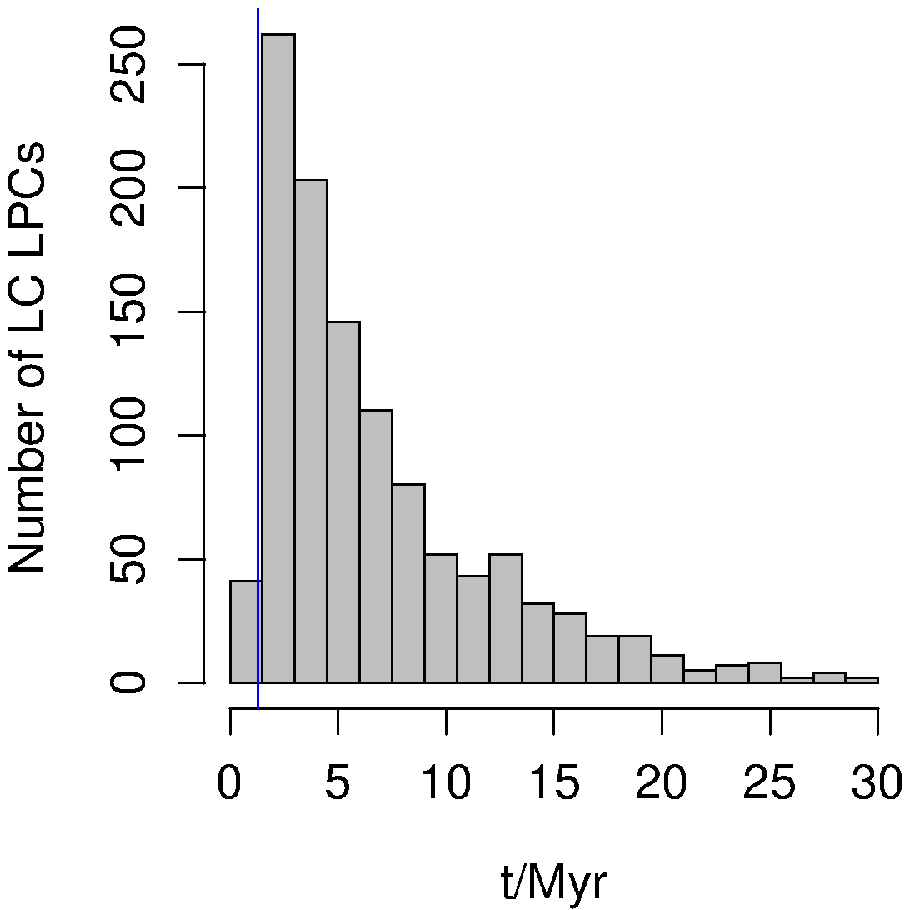}
  \includegraphics[scale=0.6]{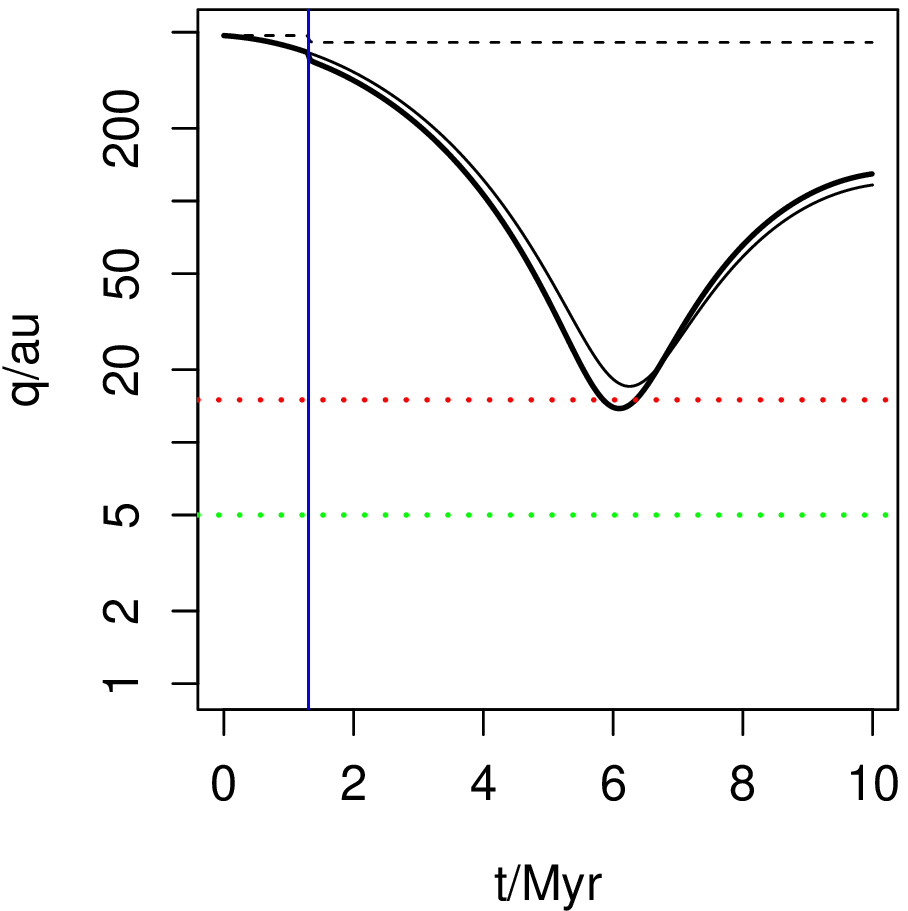}
  \caption{Upper panel: the distribution of the injection time of LPCs induced by HIP 89825 (GL 710) for a simulation of 10 million OC particles. Lower panel: the variation of the perihelion of an example LPC under the perturbations from GL 710 (dashed black line), from the tide (thin black line) and from both together (thick black line). The red and green dotted lines show the upper limits of the loss cone and observable zone respectively. The blue vertical line shows the perihelion time of GL 710 in both panels.  }
  \label{fig:89825}
\end{figure}

To illustrate further the role of encounters in perturbing the OC comets, we take HIP 89825 (GL 710) as an example. The upper panel of Figure \ref{fig:89825} shows the temporal distribution of the induced LPCs and reveals a long-term influence of HIP 89825 on the OC. In the lower panel we see that HIP 89825 can change the perihelion of an LPC instantaneously and thereby allow the tide to gradually reduce the perihelion until the comet is captured by the planets.

\begin{figure*}
  \centering
  \includegraphics[scale=0.6]{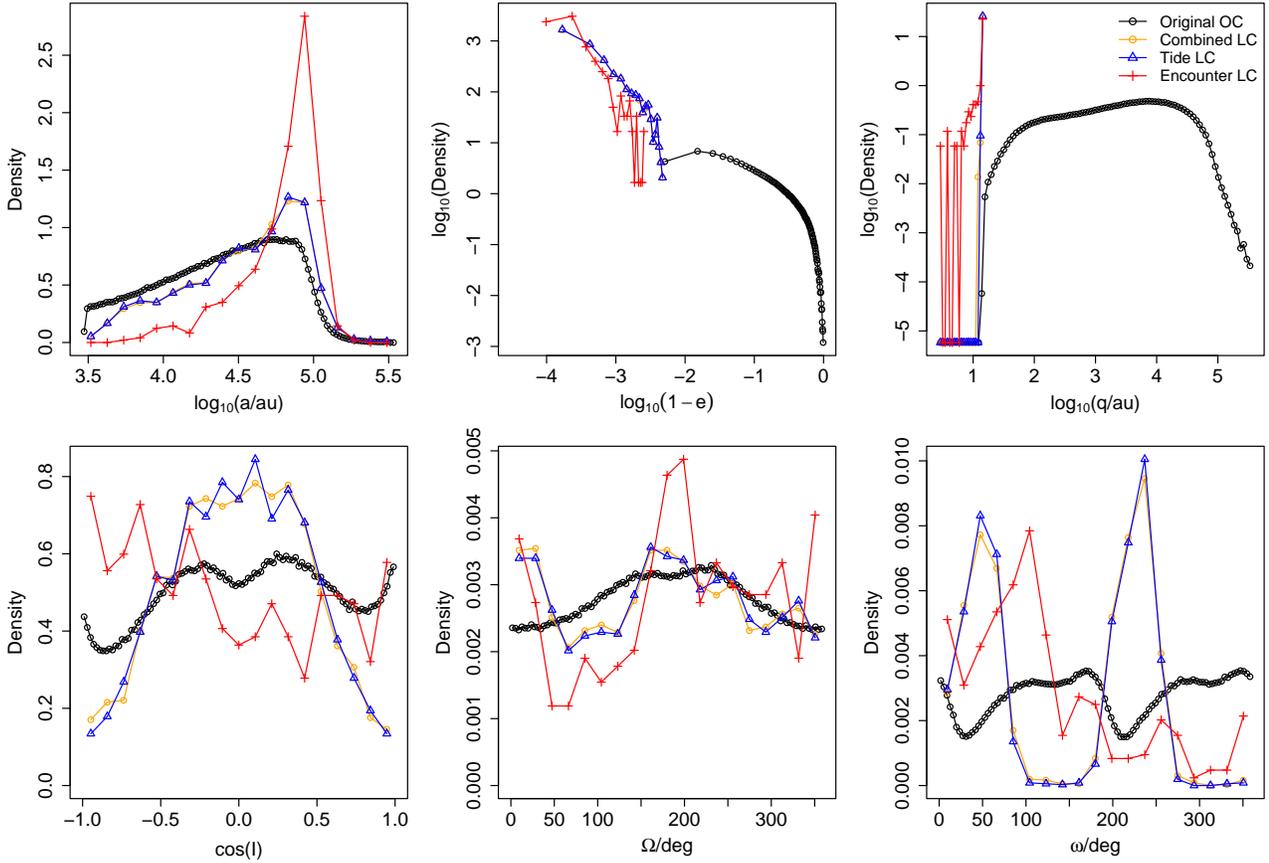}
  \caption{The distributions of orbital elements of initial OC comets (black lines) and comets that are injected into the LC in three types of simulations for HIP 103738 (other lines). The former are orbital elements of OC comets before any tidal perturbations while the latter are the ones calculated when comets are injected to become LPCs. The legend in the top right panel defines the lines for all panels: comets generated from the OC, and LPCs generated in combined, tide-only, and encounter-only simulations. In clockwise order from the top left, the panels show the: semi-major axis $a$,; eccentricity $e$; perihelion $q$; inclination $I$; longitude of ascending node $\Omega$; argument of perihelion $\omega$. Note that the distribution of $\cos I$ is uniform if the LPCs are isotropic. The number of comets in the OC and the LPCs generated in the combined, tide-only, and encounter-only simulations are $10^6$, 1892, 1912, and 444 respectively. To display the differences between encounter-induced and tide-induced LPCs better, we plot log densities over $\log_{\rm 10}(1-e)$ (to better distinguish between very high eccentricities) and 
$\log_{\rm 10}(q/{\rm au})$, rather than $e$ and $q$. The OC model used in the encounter-only simulations contains 10 million comets. The range of orbital elements are divided into 100 and 20 bins for the original OC comets and LC LPCs, respectively. The strong fluctuations of the red lines around $\log_{10}(1-e)=-3$ and for $\log_{10}(q/{\rm au})<0.8$ are caused by sample noise.}
  \label{fig:hist_element}
\end{figure*}

It is interesting also to examine the imprints of encounters on the distributions of the orbital elements of LPCs. Figure \ref{fig:hist_element} compares the orbital elements of comets in the initial OC with those of the comets generated in tide-only, encounter-only, and combined simulations for HIP 103738 (Gamma Microscopii). (For the encounter-only simulations we increased the samples size to 10 million OC comets in order to improve the statistical significance.) The orbital elements of LPCs are calculated when they are injected into the LC; this is appropriate because most orbital elements do not change significantly over the course of a single orbit. 

We performed this analysis for other encounters too. Overall we find that encounters tend to push comets with large semi-major axes into the inner solar system (i.e.\ to become LPCs), because the outer part of the OC can experience stronger tidal forces from weak encounters (with large $d_{\rm ph}$). It is necessary that all LPCs have large eccentricity and small perihelia in order to be LPCs. However, while the encounter-induced LPCs can have perihelia much less than 15\,au, tide-induced LPCs have perihelia which concentrate around the LC limit (15\,au). This highlights the important role of encounters for injecting comets into the observable zone.

In addition, tide and encounter perturbations have different influences on the other orbital elements. The tide tends to inject comets with high inclination while encounters do not. The LPCs induced by the tide generally show a W-shaped distribution in the longitude of the ascending node, $\Omega$. Finally, the argument of perihelion of tide-induced LPCs concentrate around $\omega=50$\,deg and $\omega=240$\,deg while the distribution for LPCs induced by Gamma Microscopii is concentrated around $\omega=100$\,deg. These non-uniform distributions are caused by the anisotropic perturbations imposed on the OC from the tide and encounters, which we will further discuss in section \ref{sec:angular}. 

\subsection{Comparison of perturbation strength proxies}\label{sec:proxy}

\begin{figure}
  \centering
  \includegraphics[scale=0.6]{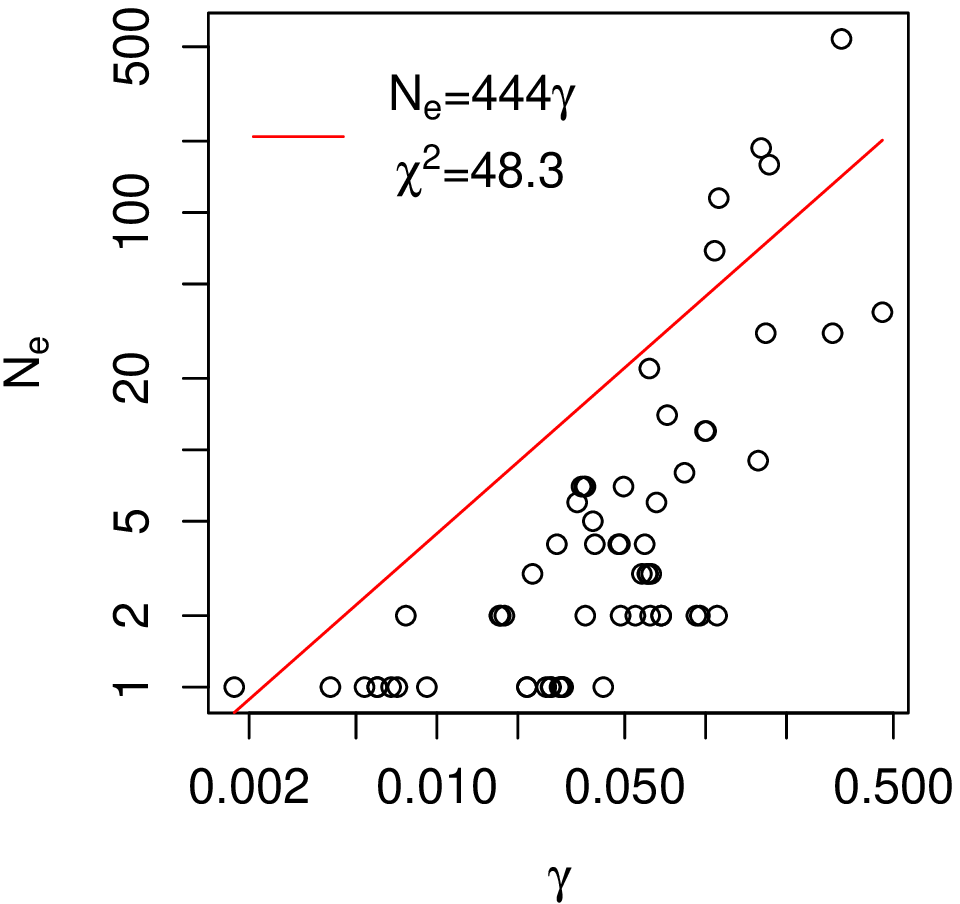}
  \includegraphics[scale=0.6]{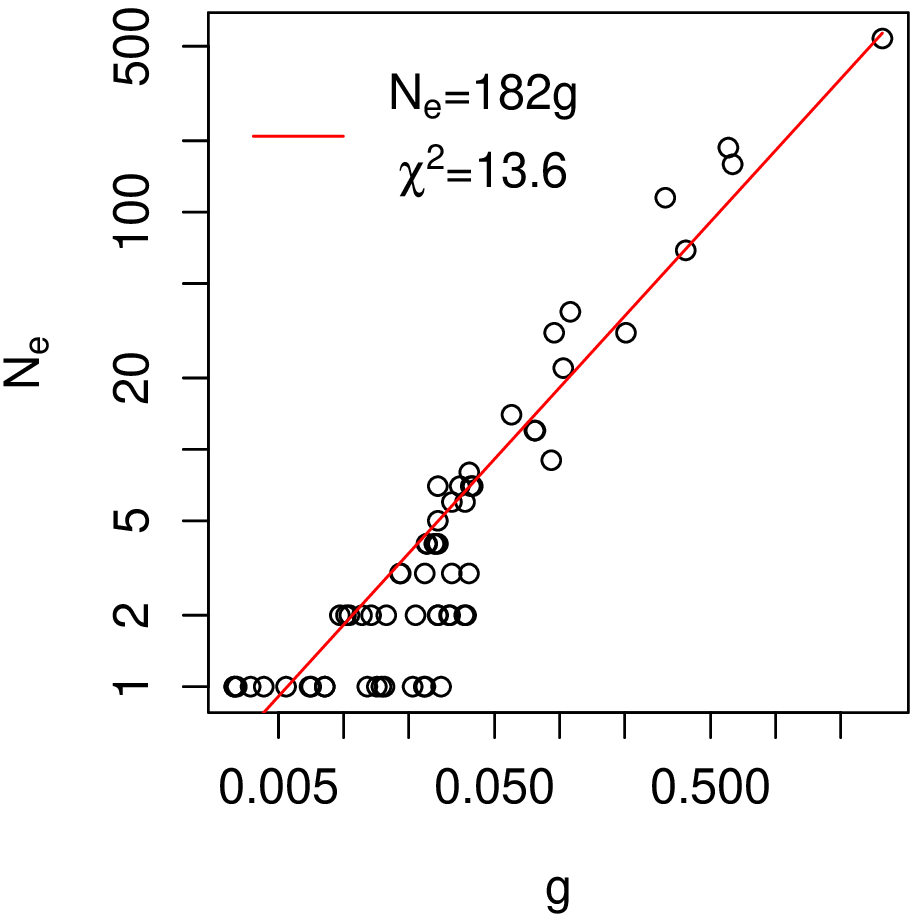}
  \caption{The relation between encounter-induced LPC flux and the two proxies,$\gamma$ (upper) and $g$ (lower). For both panels, the red line denotes the optimized fitting function. The legends show the fitted functions of $N_e(\gamma)$ and $N_e(g)$ and the Pearson's chi-square for them (with 61 degrees of freedom).}
  \label{fig:proxy}
\end{figure}

In section \ref{sec:data} we used the proxy, $g$, to select encounters. Here we compare the two proxies, $g$ and $\gamma$, with the actual number of LPCs induced, $N_e$. Figure \ref{fig:proxy} shows the log-log plot of the relation between each proxy and $N_e$. We fit linear functions of $\gamma$ and $g$ without intercept to $N_e$, i.e. $N_e=\alpha \gamma$ and $N_e=\beta g$, where $\alpha$ and $\beta$ are unknown and to be fit for. The $g$ proxy fits the number of encounter-induced LPCs with $\chi^2=12.6$ compared to $\chi^2=47.1$ for $\gamma$, indicating that $g$ is a better proxy.

We also vary the index of $d_{\rm ph}$ to find the best function form of $N_e$ which is
\begin{equation}
  N_e=\eta M_{\rm enc}/(d_{\rm ph}^\delta v_{\rm ph})\ ,
\end{equation}
where $\eta$ and $\delta$ are unknown. The best fit of the data gives an exponent of 1.92 ($\delta=1.92$) for $d_{\rm ph}$ (Eqn. \ref{eqn:proxy}), which also suggests that the $g$ proxy is better than the $\gamma$ proxy. However, this analysis only includes observed stellar encounters, which naturally includes very few rare, strong encounters. It also omits very weak encounters, which could not have perturbed any of the 1 million particles in our simulations. Thus $\gamma$ could still be a good proxy for very weak or very strong stellar encounters, as is suggested by \cite{fouchard11} and \cite{feng14}. 

\subsection{LPCs induced by encounter sequences}\label{sec:sequence}

In addition to individual encounters, we combine encounters to form encounter sequences and simulate their influence on the OC over the past and future 10\,Myr. To effectively account for the effect of weak encounters, we simulate with a sample of 10 million OC particles. We apply the tide-only, encounter-only, and combined simulations to integrate the motions of OC particles for both the reliable and the basic encounter sequences (see section \ref{sec:data}). The resulting number of LPCs is shown in Table \ref{tab:seq_LPC}. We see that over the past/future 10\,Myr, the reliable sequence of encounters contributes about 8\% to the total LPC flux (column $N_e/N_t$). The basic sequence contributes about 38\% to the total future flux, due primarily to the strong perturbations from HIP 85605. Although HIP 85605 has dubious astrometry (BJ15), we nonetheless include it in the basic sequence, not least to get an idea of the influence on the OC of possible undiscovered strong encounters. The results for the reliable sequence should be used to see the effect of known encounters, although this sequence is almost certainly incomplete.

\begin{table*}
\centering
\caption{The number of LPCs injected into the LC ($N$) and the number of dynamically new LPCs ($n$) for simulations of 10 million OC particles over the past/future 10\,Myr for the two encounter sequences. The subscripts $c$, $t$, and $e$ refer to combined, tide-only, and encounter-only respectively. The last two columns give the two synergy parameters, $s_{\rm lc}$ and $s_{\rm new}$.}
\label{tab:seq_LPC}
\begin{tabular}{c|*9{c|}c|c}
  \hline
  Sequence&Time span&$N_c$&$N_t$&$N_e$&$n_c$&$n_t$&$n_e$&$N_e/N_t$&$n_e/n_t$&$s_{\rm lc}$(\%)&$s_{\rm new}$(\%)\\
  \hline
  \multirow{ 2}{*}{Reliable }&-10 to 0\,Myr &19735&18672&1571&7089&6634&281&0.084&0.042&-2.3&2.5\\
  &0 to 10\,Myr&20151&18873&1560&7147&6708&200&0.083&0.030&-1.4&3.3\\\hline
  \multirow{ 2}{*}{Basic }&-10 to 0\,Myr&19764&18672&1604&7125&6634&279&0.086&0.042&-2.6&3.0\\
  &0 to 10\,Myr&24571&18873&7080&8441&6708&1525&0.38&0.23&-5.6&2.5\\
  \hline
\end{tabular}
\end{table*}

Table \ref{tab:seq_LPC} also shows that encounters seem less able to generate dynamically new comets by themselves (column $n_e/n_t$). This is because they cannot change the perihelion sufficiently over a short interaction period (typically a few thousand years). The tide, in contrast, can gradually reduce the perihelia of comets by a large amount. Assuming there are $10^{12}$ comets in the OC \citep{weissman90nat,weissman96,dones04a}, the averaged rate of dynamically new LPCs (with perihelia less than 5\,au) is about 70 per year according to our simulations. For combined simulations for the reliable sequence, there are about 30 new comets per year which approach within 1\,au of the Sun. This is about one order of magnitude larger than the discovery rate of 2.1 new comets (with $q<1$\,au) per year \citep{dones04a}. If the DLDW model is correct and the planets do not change the flux of new LPC over a 10\,Myr time span, and if most dynamically new comets with $q<1$\,au are indeed discovered, then our simulations suggest a smaller population of OC comets, of order $10^{11}$ comets, and thus a smaller total mass of the OC.  

We can also investigate the nonlinear effect caused by the combined perturbations from the tide and encounters (the so-called ``synergy effect'') in delivering LPCs from the OC \citep{rickman08}. The synergy effect is measured by the fractional difference in number of LPCs injected into the LC by the combined vs.\ encounter and tide alone, i.e.\
\begin{equation}
  s_{\rm lc} = (N_c - N_e - N_t)/N_c .
\end{equation}
The synergy parameter for just the dynamically new comets is similarly
\begin{equation}
  s_{\rm new} = (n_c - n_e - n_t)/n_c .
\end{equation}
Values of $s$ for both sequences and time spans are shown in Table \ref{tab:seq_LPC}. The synergy effect is not significant -- only a few percent -- for both sequences for both dynamically new and LC LPCs. This is small compared to the synergy effect found in the late stage of the 5\,Gyr simulations for simulated encounters performed by \cite{rickman08}. This is probably a consequence of the relatively short time span of our simulations which does not permit us to include the long-term effect of massive encounters, plus the fact that we have very few massive encounters in our real data set \citep{fouchard11}.

Generally, the synergy effect is negative for LC LPCs, but is positive for dynamically new LPCs. An encounter may perturb OC comets in the opposite direction to the tide, thus reducing the LC LPC flux compared to when either acts alone. Thus the negative synergy effect for the sequence may be caused by one or a few strong encounters. In contrast, the tide is not sufficient to inject comets into the observable zone directly (see the top right panel of Figure \ref{fig:hist_element}). Thus new LPCs tend to be injected by the combined impulse from the tide and encounters, leading to a positive synergy effect.  

\begin{figure}
\hspace{-0.2in}
  \begin{minipage}{1.\linewidth}
  \includegraphics[scale=0.45]{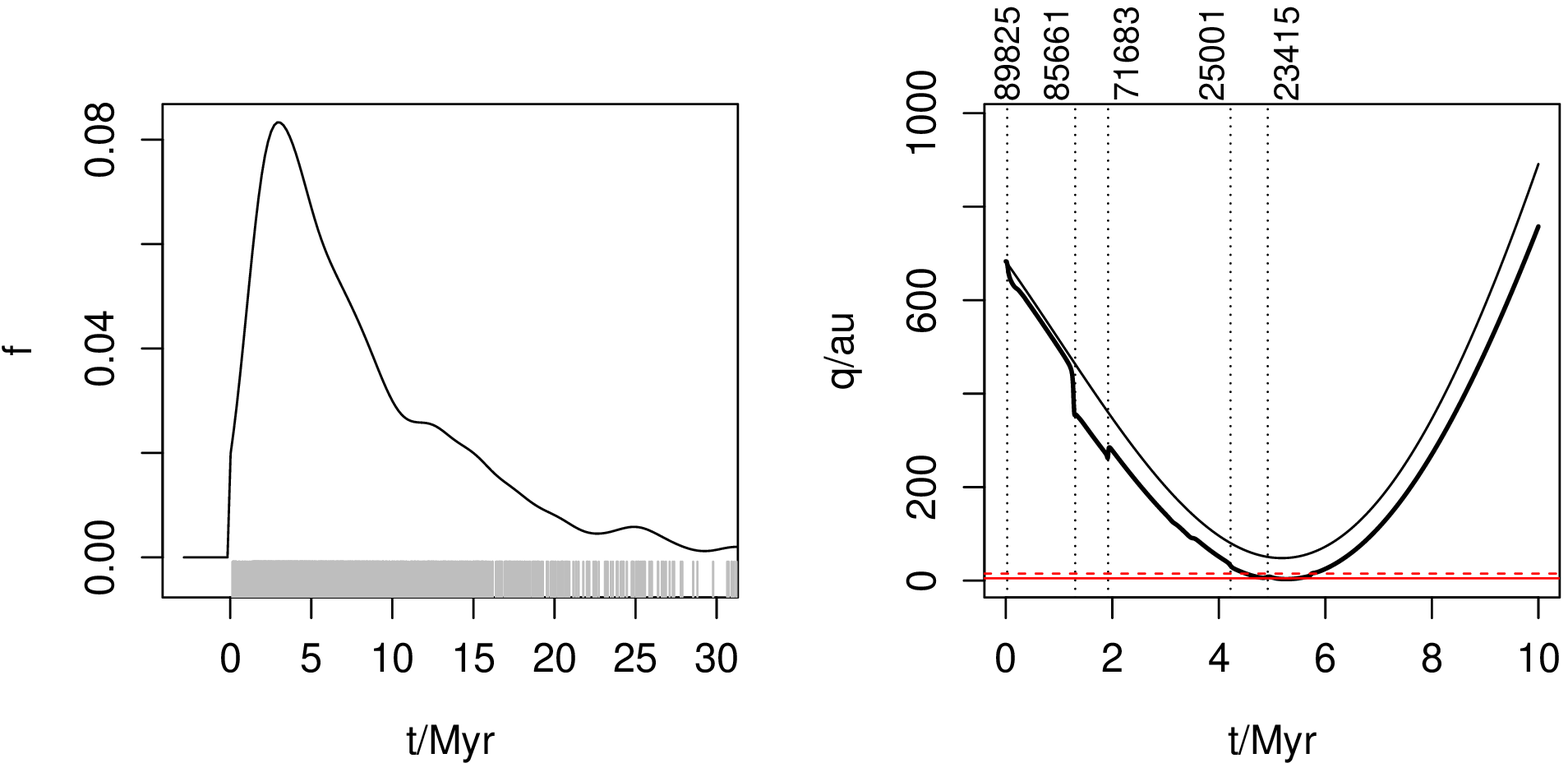}
  \includegraphics[scale=0.45]{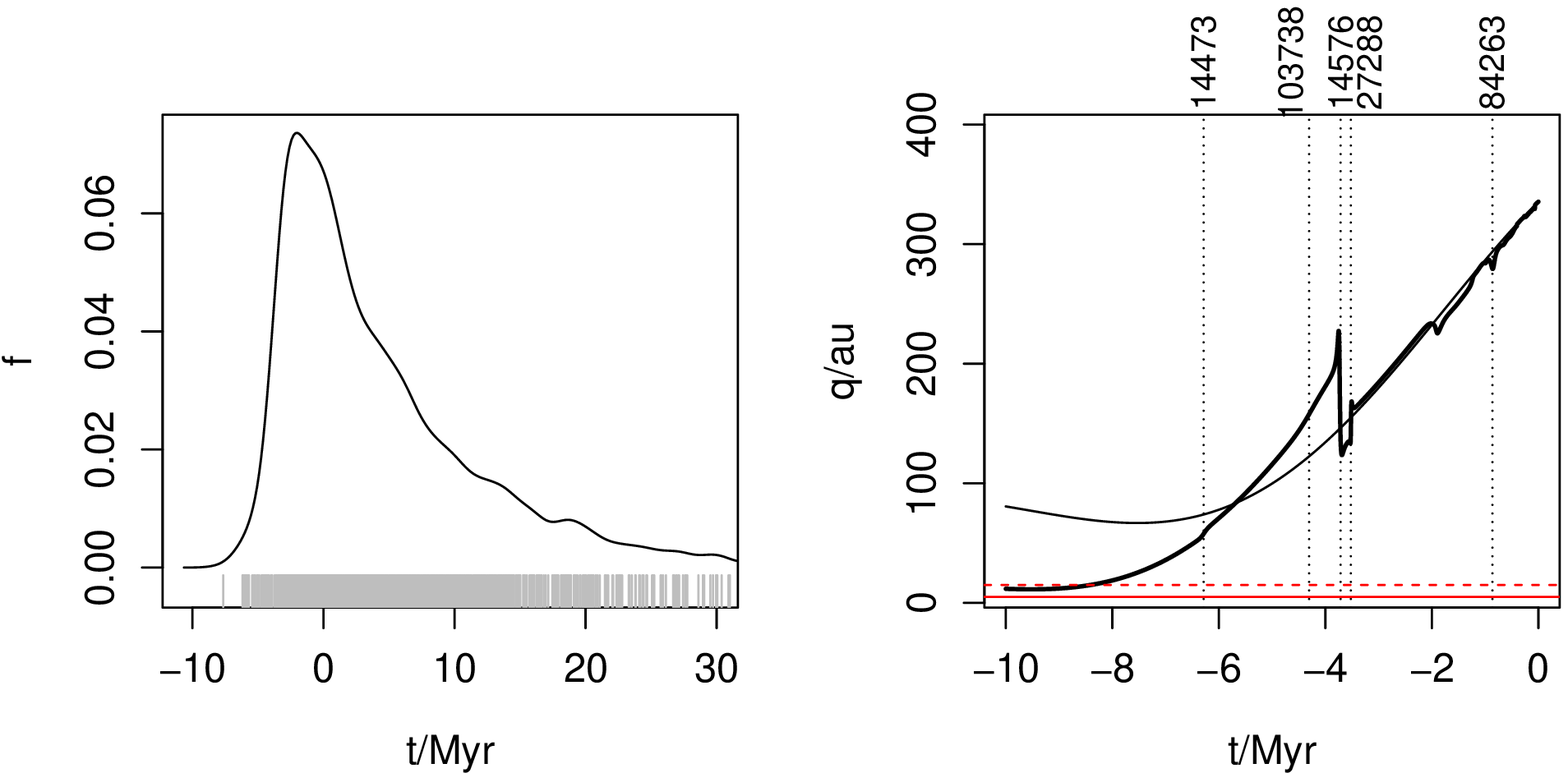}
  \end{minipage}
  \caption{The flux ratio, $f$, as a function of time (left panels) and the change of the perihelion of an example LPC (right panels) for the reliable sequence for future (top) and past (bottom) encounters. The left panels are similar to Figure \ref{fig:density}. Note that $f(t<0)$ is set to be zero for future encounters. The right panels show the perihelion variation of an example LPC for both the tide-only (thin) and the combined perturbations (thick). The perihelion times of the five strongest future/past encounters are shown with vertical dotted lines and denoted by their Hipparcos IDs.}
  \label{fig:reliable_distribution}
\end{figure}

We may also examine the flux ratio of encounter-induced LPCs over time and the evolution of the perihelia of some example LPCs. This is shown in Figure \ref{fig:reliable_distribution} for the reliable sequence. The shape of the flux ratio in the left two panels is driven by the observational selection effects in our sample: there are fewer encounters the further we go into the past or future and so fewer injections at times far from the present. Nonetheless, injections can occur well after the encounter perihelion.
According to the left two panels, the contribution of past encounters to the current LPC flux is about 6\%, which is only a lower limit due to the incompleteness. The past-future asymmetry in the flux ratio and the sharp decrease around $t=0$\,Myr are caused by the time required by LPCs to reach their perihelia. Most encounters have a perihelion time within 5\,Myr from now, and have orbital period of around 5\,Myr. Thus the flux ratio reaches its maximum at around 4\,Myr and -1\,Myr for the future and past sequences of encounters respectively. We see in the right panels that encounters can always intensify the variation of the perihelion of an LPC, and thus make it more likely to enter into the observable zone from outside of the LC within one revolution. This again explains why the synergy effect is positive for new LPCs. 

\vspace{-0.2in}

\subsection{Angular distribution of the perihelia of LPCs and encounters}\label{sec:angular}

Many previous studies found that the dynamically new LPCs have anisotropic perihelia (\citealt{delsemme87,matese11}; FBJ14). Here we reexamine this by investigating the effect of encounters on the angular distribution of the LPC perihelia.

\begin{figure*}
  \centering
  \includegraphics[scale=0.49]{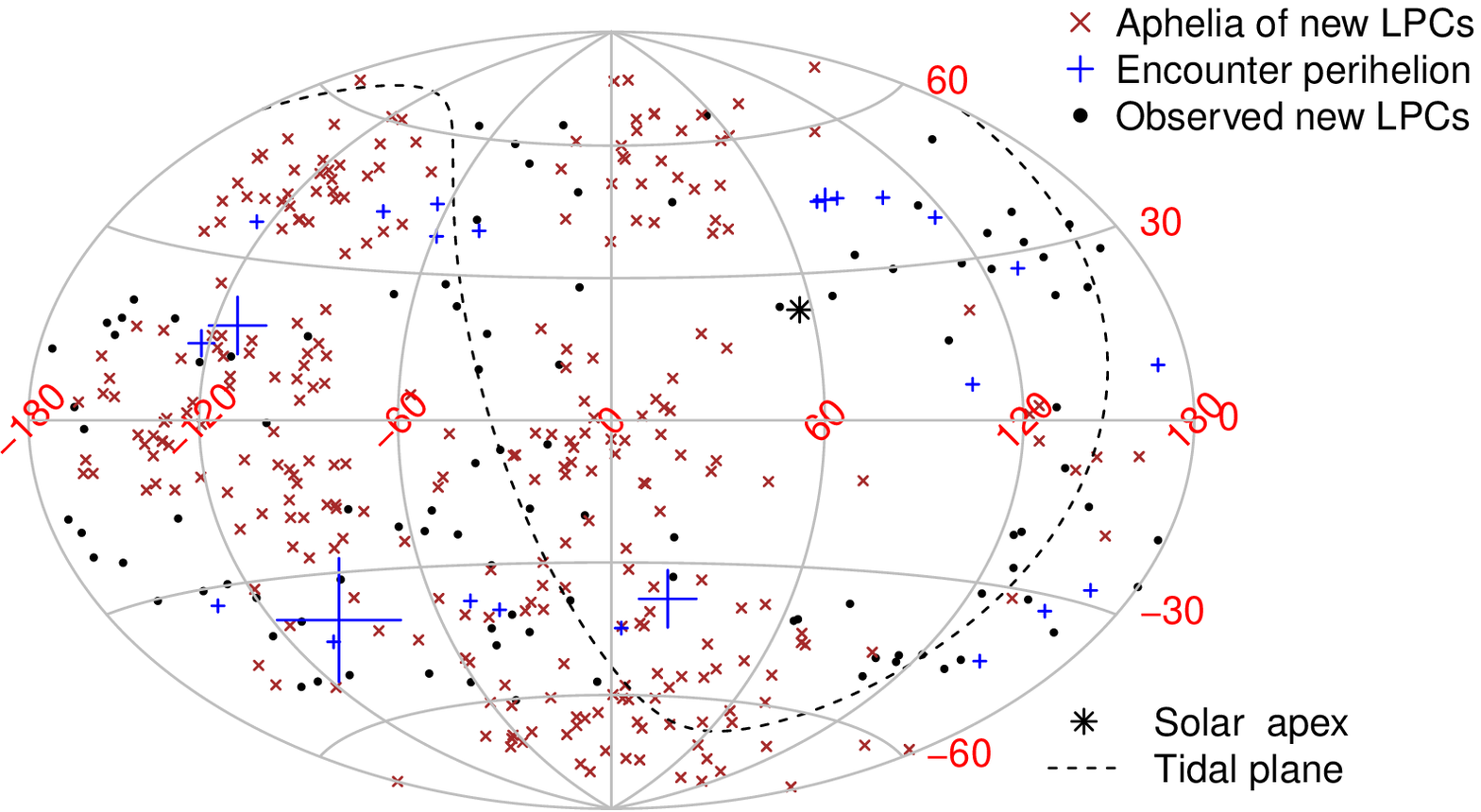}
  \includegraphics[scale=0.49]{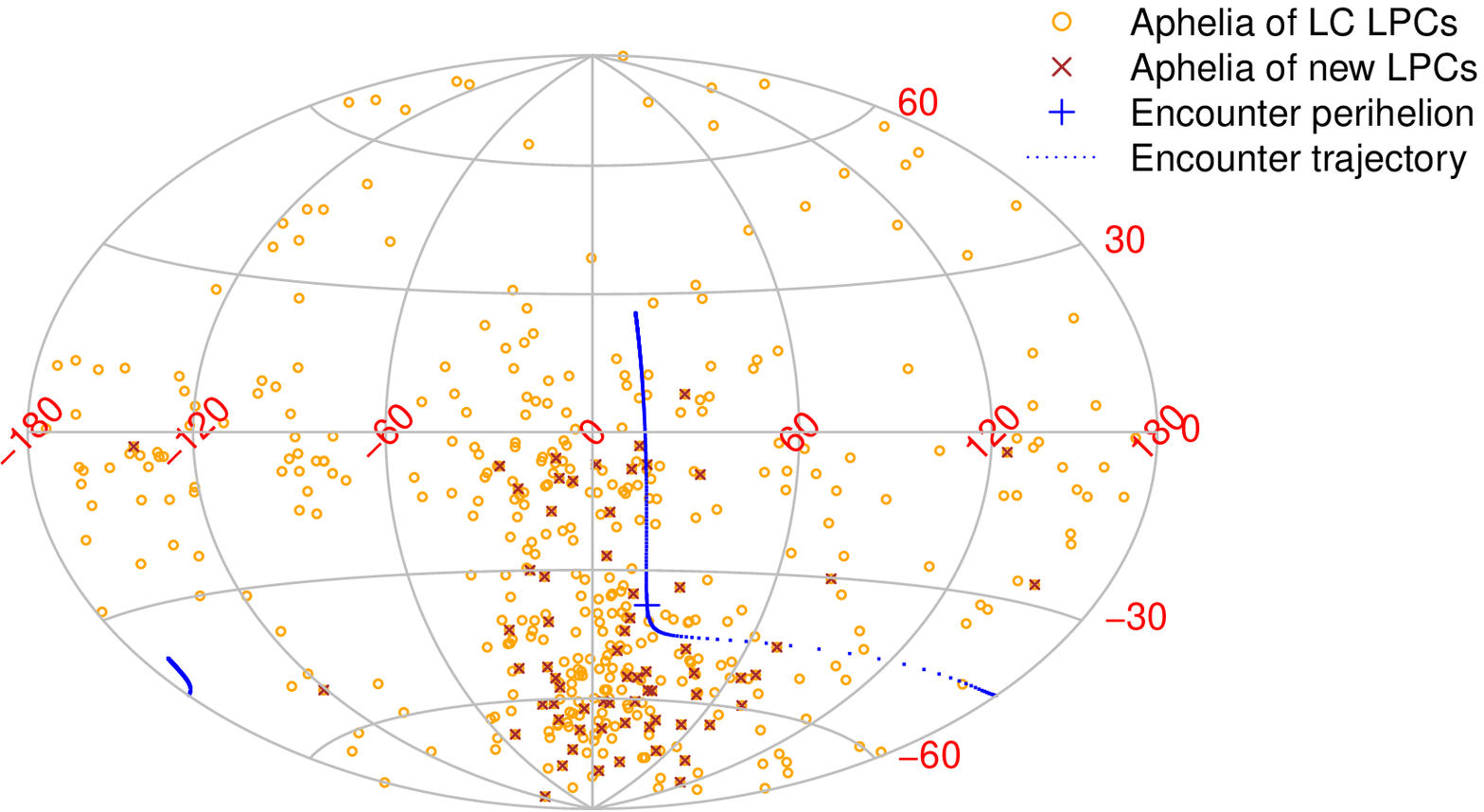}
  \caption{An Aitoff projection in Galactic coordinates of the aphelia/perihelia of LPCs/encounters for the reliable sequence of past encounters (left panel) and HIP 103738 (Gamma Microscopii, right panel). In the left panel, the blue crosses show the perihelion of the encounters. The size of the cross is proportional to the number of LPCs the encounter has induced according to Table \ref{tab:encounter} (although a lower limit of the size is set for very weak encounters). The red crosses show the aphelia of dynamically new LPCs in the simulation and the black points show the aphelia of observed dynamically new LPCs.
The black star symbol indicates the solar apex, and the dashed line shows the preferred tidal plane, the normal to which is the solar apex (FBJ14). The right panel is similar, but is only for HIP 103738, and does not show the observed LPCs and additionally shows the aphelia of LC LPCs in the simulation (yellow open circles). The blue curve is actually a sequence of points showing the encounter trajectory when its pericentric distance is less than 4\,pc. Each point gives the star's position at a equally-separated time interval of 0.001\,Myr. 
}
  \label{fig:aitoff}
\end{figure*}

The left panel of Figure \ref{fig:aitoff} shows, in Galactic coordinates, the positions of the aphelia of encounter-induced LPCs (red crosses) and the perihelia of the stellar encounters (blue crosses) for the reliable sequence of past encounters.  We observe an East-West asymmetry in the directions of the aphelia of LPCs. This is due to the inhomogeneity in the distribution of strong encounters. We also see a significant difference between the locations of the aphelia of the observed LPCs (black points) and the simulated LPCs (red crosses). We further investigated the aphelia of LPCs induced by the tide and find a similar discrepancy.  Given that the set of encounters we have is incomplete, and only accounts for about 5\% of the LPC flux, it is not surprising that we see little correlation. The observed set of LPCs is also likely to be incomplete (as what we see ``now'' only covers a few tens of years).

We also see that the observed LPCs' aphelia are located around the preferred tidal plane, where perturbations imposed on the OC are stronger due to the solar apex motion (FBJ14). In contrast, the perihelia of encounters are only weakly associated with the tidal plane.  A larger sample of encounters is required to confirm the existence of the tidal plane in order to reduce the sample noise. We get similar results when we analyze the LPCs induced by the reliable sequence of future encounters.

The right panel of Figure \ref{fig:aitoff} shows the aphelia of LPCs induced just by HIP 103738 (Gamma Microscopii), one of the strongest recent past encounters. We find that the aphelia of both LC LPCs and dynamically new LPCs cluster near the trajectory of the encounter, although the geometry of the OC and characteristics of encounters complicates the spatial distribution of LPCs' perihelia. In particular, the aphelia of LPCs tend to concentrate around the location where the encounter was close to its perihelion. This is caused by the fact that an encounter generally has stronger geometric preference in imposing perturbations on the OC when it approaches its perihelion.

\section{Discussion and conclusion}\label{sec:conclusions}

We have conducted simulations of the perturbation of the Oort cloud in order to estimate the significance of known encounters in generating long period comets. According to simulations both of individual encounters and encounter sequences, the past encounters in our sample contribute about 5\% of the current observed LPCs. Most LPCs were instead injected from the OC by the Galactic tide. We find that HIP 89825 (GL 710) and HIP 14473 are the strongest future and past encounters in our sample respectively, when we exclude objects with questionable data. 
We have also quantified the effect of other strong encounters. For most encounters, the peak of the encounter-induced LPC flux occurs about 2\,Myr after the time of their perihelia.

Consistent with the analytical estimations of the strength of encounters given by \cite{fouchard11} and \cite{matese02}, our simulations show that the proxy $g=M_{\rm enc}/(d_{\rm ph}^2 v_{\rm ph})$ can approximate the LPC flux induced by medium encounters reasonably well. Although some stars came (or will come) very close to the solar system, massive stars generally play a much larger role in perturbing the OC and increasing the flux of LPCs. In particular, the recently noted close encounter by W0720 \cite{mamajek15} was extremely weak.

We find that the synergy effect between encounters and the Galactic tide is not significant over 10\,Myr time scales, although it could be significant over longer time scales. A comparison of the simulated and observed flux of new LPCs suggests a population size of only $10^{11}$ comets in the Oort cloud, which would require a smaller initial mass of the solar nebula than is often assumed. This may resolve a possible inconsistency between the OC mass and planet formation theories \citep{morbidelli05,duncan08}. We also find that the perihelia of encounter-induced LPCs generally follow the trajectories of strong encounters. Since the tide cannot fully explain the anisotropic perihelia of LPCs, we expect that there are strong stellar encounters which have not yet been discovered. These may well be found by future astrometric surveys.

Without accounting for the perturbations from the planets and observation bias, our model cannot precisely predict the current LPC flux. Yet our results on the flux of just dynamically new LPCs are more reliable because new LPCs have enough energy to penetrate the Jupiter-Saturn barrier \citep{heisler86}. Although the new LPCs can also form through the Kaib-Quinn mechanism \citep{kaib09}, the DLDW model has already taken this into account, and the planets can barely influence the flux of new LPCs over a 10\,Myr time span.

The catalogs of encounters we have used are mainly derived from Hipparcos, limiting us to stars with $V<12$ and with significant incompleteness. The Hipparcos catalogue has probably missed about one half of the encounters \citep{feng15dis}, which have encountered (or will encounter) the solar system over the past (or future) 10\,Myr. Gaia \citep{perryman01} should detect more than 90\% encounters over the past/future 10\,Myr and provide accurate astrometric and spectrophotometric measurements for them \citep{feng15dis}. Our work shows that future studies of encounters should focus not only on close encounters but also on massive and slow encounters.

\section*{Acknowledgements}

We used the ViziR catalogue access tool and the Simbad object database at CDS to collect data on the encounters. We thank Piotr Dybczy\'nski and Filip Berski for providing the data they used to simulate stellar encounters. Morgan Fouesneau, Ren\'e Andrae, and Kester Smith gave valuable comments on the manuscript. Liang Wang, Dae-Won Kim, and Xiangxiang Xue gave technical support for the revision of the manuscript. 

\bibliographystyle{mn2e}
\bibliography{comet.bib}

\begin{thebibliography}{61}
\expandafter\ifx\csname natexlab\endcsname\relax\def\natexlab#1{#1}\fi

\bibitem[{{Alvarez} {et~al}\mbox{.}(1980){Alvarez}, {Alvarez}, {Asaro}, \&
  {Michel}}]{alvarez80}
{Alvarez} L.~W., {Alvarez} W., {Asaro} F., {Michel} H.~V., 1980, Science, 208,
  1095

\bibitem[{{Anderson} \& {Francis}(2012)}]{anderson12}
{Anderson} E., {Francis} C., 2012, Astronomy Letters, 38, 331

\bibitem[{{Bailer-Jones}(2015)}]{bailer-jones15}
{Bailer-Jones} C.~A.~L., 2015, \aap, 575, A35 (BJ15)

\bibitem[{{Bobylev}(2010)}]{bobylev10b}
{Bobylev} V.~V., 2010, Astronomy Letters, 36, 220

\bibitem[{Bottke {et~al}\mbox{.}(2005)Bottke, Durda, Nesvorn{\`y}, Jedicke,
  Morbidelli, Vokrouhlick{\`y}, \& Levison}]{bottke05}
Bottke W.~F., Durda D.~D., Nesvorn{\`y} D., Jedicke R., Morbidelli A.,
  Vokrouhlick{\`y} D., Levison H.~F., 2005, Icarus, 179, 63

\bibitem[{Brasser {et~al}\mbox{.}(2012)Brasser, Duncan, Levison, Schwamb, \&
  Brown}]{brasser12}
Brasser R., Duncan M., Levison H., Schwamb M., Brown M., 2012, Icarus, 217, 1

\bibitem[{{Brown}, {Trujillo} \& {Rabinowitz}(2004){Brown}, {Trujillo}, \&
  {Rabinowitz}}]{brown04}
{Brown} M.~E., {Trujillo} C., {Rabinowitz} D., 2004, \apj, 617, 645

\bibitem[{{Casagrande} {et~al}\mbox{.}(2011){Casagrande}, {Sch{\"o}nrich},
  {Asplund}, {Cassisi}, {Ram{\'{\i}}rez}, {Mel{\'e}ndez}, {Bensby}, \&
  {Feltzing}}]{casagrande11}
{Casagrande} L., {Sch{\"o}nrich} R., {Asplund} M., {Cassisi} S.,
  {Ram{\'{\i}}rez} I., {Mel{\'e}ndez} J., {Bensby} T., {Feltzing} S., 2011,
  \aap, 530, A138

\bibitem[{Culler {et~al}\mbox{.}(2000)Culler, Becker, Muller, \&
  Renne}]{culler00}
Culler T.~S., Becker T.~A., Muller R.~A., Renne P.~R., 2000, Science, 287, 1785

\bibitem[{{De Rosa} {et~al}\mbox{.}(2012){De Rosa}, {Patience}, {Vigan},
  {Wilson}, {Schneider}, {McConnell}, {Wiktorowicz}, {Marois}, {Song},
  {Macintosh}, {Graham}, {Bessell}, {Doyon}, \& {Lai}}]{derosa12}
{De Rosa} R.~J. {et~al.}, 2012, \mnras, 422, 2765

\bibitem[{{Delsemme}(1987)}]{delsemme87}
{Delsemme} A.~H., 1987, \aap, 187, 913

\bibitem[{{Dones} {et~al}\mbox{.}(2004{\natexlab{a}}){Dones}, {Levison},
  {Duncan}, \& {Weissman}}]{dones04b}
{Dones} L., {Levison} H.~F., {Duncan} M.~J., {Weissman} P.~R.,
  2004{\natexlab{a}}, Icarus, in press

\bibitem[{{Dones} {et~al}\mbox{.}(2004{\natexlab{b}}){Dones}, {Weissman},
  {Levison}, \& {Duncan}}]{dones04a}
{Dones} L., {Weissman} P.~R., {Levison} H.~F., {Duncan} M.~J.,
  2004{\natexlab{b}}, {Oort cloud formation and dynamics}, {Festou} M.~C.,
  {Keller} H.~U., {Weaver} H.~A., eds., pp. 153--174

\bibitem[{{Duncan} {et~al}\mbox{.}(2008){Duncan}, {Brasser}, {Dones}, \&
  {Levison}}]{duncan08}
{Duncan} M.~J., {Brasser} R., {Dones} L., {Levison} H.~F., 2008, {The Role of
  the Galaxy in the Dynamical Evolution of Transneptunian Objects}, {Barucci}
  M.~A., {Boehnhardt} H., {Cruikshank} D.~P., {Morbidelli} A., {Dotson} R.,
  eds., pp. 315--331

\bibitem[{{Dybczy{\'n}ski}(2002)}]{dybczynski02}
{Dybczy{\'n}ski} P.~A., 2002, \aap, 396, 283

\bibitem[{{Dybczy{\'n}ski} \& {Berski}(2015)}]{dybczynski15b}
{Dybczy{\'n}ski} P.~A., {Berski} F., 2015, \mnras, 449, 2459

\bibitem[{{Dybczy{\'n}ski} \& {Kr{\'o}likowska}(2015)}]{dybczynski15a}
{Dybczy{\'n}ski} P.~A., {Kr{\'o}likowska} M., 2015, \mnras, 448, 588

\bibitem[{Everhart(1967)}]{everhart67}
Everhart E., 1967, \apj, 72, 716

\bibitem[{{Feng}(2015)}]{feng15dis}
{Feng} F., 2015, PhD thesis, preprint (arXiv:1505.07856)

\bibitem[{{Feng} \& {Bailer-Jones}(2013)}]{feng13}
{Feng} F., {Bailer-Jones} C.~A.~L., 2013, \apj, 768, 152

\bibitem[{Feng \& Bailer-Jones(2014)}]{feng14}
Feng F., Bailer-Jones C. A.~L., 2014, \mnras, 442, 3653 (FBJ14)

\bibitem[{{Fouchard} {et~al}\mbox{.}(2011){Fouchard}, {Froeschl{\'e}},
  {Rickman}, \& {Valsecchi}}]{fouchard11}
{Fouchard} M., {Froeschl{\'e}} C., {Rickman} H., {Valsecchi} G.~B., 2011,
  \icarus, 214, 334

\bibitem[{{Garc{\'{\i}}a-S{\'a}nchez}
  {et~al}\mbox{.}(1999){Garc{\'{\i}}a-S{\'a}nchez}, {Preston}, {Jones},
  {Weissman}, {Lestrade}, {Latham}, \& {Stefanik}}]{sanchez99}
{Garc{\'{\i}}a-S{\'a}nchez} J., {Preston} R.~A., {Jones} D.~L., {Weissman}
  P.~R., {Lestrade} J.-F., {Latham} D.~W., {Stefanik} R.~P., 1999, \aj, 117,
  1042

\bibitem[{{Gardner} {et~al}\mbox{.}(2011){Gardner}, {Nurmi}, {Flynn}, \&
  {Mikkola}}]{gardner11}
{Gardner} E., {Nurmi} P., {Flynn} C., {Mikkola} S., 2011, \mnras, 411, 947

\bibitem[{Gomes {et~al}\mbox{.}(2005)Gomes, Levison, Tsiganis, \&
  Morbidelli}]{gomes05}
Gomes R., Levison H.~F., Tsiganis K., Morbidelli A., 2005, Nature, 435, 466

\bibitem[{{Gontcharov}(2006)}]{gontcharov06}
{Gontcharov} G.~A., 2006, Astronomy Letters, 32, 759

\bibitem[{{Heisler} \& {Tremaine}(1986)}]{heisler86}
{Heisler} J., {Tremaine} S., 1986, \icarus, 65, 13

\bibitem[{{Heisler}, {Tremaine} \& {Alcock}(1987){Heisler}, {Tremaine}, \&
  {Alcock}}]{heisler87}
{Heisler} J., {Tremaine} S., {Alcock} C., 1987, \icarus, 70, 269

\bibitem[{Ivanov(2002)}]{ivanov02}
Ivanov B., 2002, Asteroids III, 1, 89

\bibitem[{Jourdan, Reimold \& Deutsch(2012)Jourdan, Reimold, \&
  Deutsch}]{jourdan12}
Jourdan F., Reimold W.~U., Deutsch A., 2012, Elements, 8, 49

\bibitem[{{Kaib} \& {Quinn}(2009)}]{kaib09}
{Kaib} N.~A., {Quinn} T., 2009, Science, 325, 1234

\bibitem[{{Kaib}, {Ro{\v s}kar} \& {Quinn}(2011){Kaib}, {Ro{\v s}kar}, \&
  {Quinn}}]{kaib11}
{Kaib} N.~A., {Ro{\v s}kar} R., {Quinn} T., 2011, \icarus, 215, 491

\bibitem[{{Kenyon} \& {Bromley}(2004)}]{kenyon04}
{Kenyon} S.~J., {Bromley} B.~C., 2004, \nat, 432, 598

\bibitem[{{Kordopatis} {et~al}\mbox{.}(2013){Kordopatis}, {Gilmore},
  {Steinmetz}, {Boeche}, {Seabroke}, {Siebert}, {Zwitter}, {Binney}, {de
  Laverny}, {Recio-Blanco}, {Williams}, {Piffl}, {Enke}, {Roeser}, {Bijaoui},
  {Wyse}, {Freeman}, {Munari}, {Carrillo}, {Anguiano}, {Burton}, {Campbell},
  {Cass}, {Fiegert}, {Hartley}, {Parker}, {Reid}, {Ritter}, {Russell},
  {Stupar}, {Watson}, {Bienaym{\'e}}, {Bland-Hawthorn}, {Gerhard}, {Gibson},
  {Grebel}, {Helmi}, {Navarro}, {Conrad}, {Famaey}, {Faure}, {Just}, {Kos},
  {Matijevi{\v c}}, {McMillan}, {Minchev}, {Scholz}, {Sharma}, {Siviero}, {de
  Boer}, \& {{\v Z}erjal}}]{kordopatis13}
{Kordopatis} G. {et~al.}, 2013, \aj, 146, 134

\bibitem[{{Levison} {et~al}\mbox{.}(2010){Levison}, {Duncan}, {Brasser}, \&
  {Kaufmann}}]{levison10}
{Levison} H.~F., {Duncan} M.~J., {Brasser} R., {Kaufmann} D.~E., 2010, Science,
  329, 187

\bibitem[{{Liebert} {et~al}\mbox{.}(2005){Liebert}, {Young}, {Arnett},
  {Holberg}, \& {Williams}}]{liebert05}
{Liebert} J., {Young} P.~A., {Arnett} D., {Holberg} J.~B., {Williams} K.~A.,
  2005, \apjl, 630, L69

\bibitem[{{Malagnini} \& {Morossi}(1990)}]{malagnini90}
{Malagnini} M.~L., {Morossi} C., 1990, \aaps, 85, 1015

\bibitem[{{Mamajek} {et~al}\mbox{.}(2015){Mamajek}, {Barenfeld}, {Ivanov},
  {Kniazev}, {V{\"a}is{\"a}nen}, {Beletsky}, \& {Boffin}}]{mamajek15}
{Mamajek} E.~E., {Barenfeld} S.~A., {Ivanov} V.~D., {Kniazev} A.~Y.,
  {V{\"a}is{\"a}nen} P., {Beletsky} Y., {Boffin} H.~M.~J., 2015, \apjl, 800,
  L17

\bibitem[{{Matese} \& {Lissauer}(2002)}]{matese02}
{Matese} J.~J., {Lissauer} J.~J., 2002, \icarus, 157, 228

\bibitem[{{Matese} \& {Whitmire}(2011)}]{matese11}
{Matese} J.~J., {Whitmire} D.~P., 2011, \icarus, 211, 926

\bibitem[{{Morbidelli}(2005)}]{morbidelli05}
{Morbidelli} A., 2005, ArXiv Astrophysics e-prints

\bibitem[{{Morbidelli} \& {Levison}(2004)}]{morbidelli04}
{Morbidelli} A., {Levison} H.~F., 2004, \aj, 128, 2564

\bibitem[{{Oort}(1950)}]{oort50}
{Oort} J.~H., 1950, \bain, 11, 91

\bibitem[{{Pelupessy}, {J{\"a}nes} \& {Portegies Zwart}(2012){Pelupessy},
  {J{\"a}nes}, \& {Portegies Zwart}}]{pelupessy12}
{Pelupessy} F.~I., {J{\"a}nes} J., {Portegies Zwart} S., 2012, \na, 17, 711

\bibitem[{{Perryman} {et~al}\mbox{.}(2001){Perryman}, {de Boer}, {Gilmore},
  {H{\o}g}, {Lattanzi}, {Lindegren}, {Luri}, {Mignard}, {Pace}, \& {de
  Zeeuw}}]{perryman01}
{Perryman} M.~A.~C. {et~al.}, 2001, \aap, 369, 339

\bibitem[{{Portegies Zwart} {et~al}\mbox{.}(2009){Portegies Zwart}, {McMillan},
  {Harfst}, {Groen}, {Fujii}, {Nuall{\'a}in}, {Glebbeek}, {Heggie}, {Lombardi},
  {Hut}, {Angelou}, {Banerjee}, {Belkus}, {Fragos}, {Fregeau}, {Gaburov},
  {Izzard}, {Juri{\'c}}, {Justham}, {Sottoriva}, {Teuben}, {van Bever},
  {Yaron}, \& {Zemp}}]{portegies09}
{Portegies Zwart} S. {et~al.}, 2009, \na, 14, 369

\bibitem[{{Rhee} {et~al}\mbox{.}(2007){Rhee}, {Song}, {Zuckerman}, \&
  {McElwain}}]{rhee07}
{Rhee} J.~H., {Song} I., {Zuckerman} B., {McElwain} M., 2007, \apj, 660, 1556

\bibitem[{{Rickman}(1976)}]{rickman76}
{Rickman} H., 1976, Bulletin of the Astronomical Institutes of Czechoslovakia,
  27, 92

\bibitem[{{Rickman} {et~al}\mbox{.}(2008){Rickman}, {Fouchard},
  {Froeschl{\'e}}, \& {Valsecchi}}]{rickman08}
{Rickman} H., {Fouchard} M., {Froeschl{\'e}} C., {Valsecchi} G.~B., 2008,
  Celestial Mechanics and Dynamical Astronomy, 102, 111

\bibitem[{{Rickman} {et~al}\mbox{.}(2012){Rickman}, {Fouchard},
  {Froeschl{\'e}}, \& {Valsecchi}}]{rickman12}
{Rickman} H., {Fouchard} M., {Froeschl{\'e}} C., {Valsecchi} G.~B., 2012,
  \planss, 73, 124

\bibitem[{Scholz, Bihain \& Storm(2014)Scholz, Bihain, \& Storm}]{scholz14}
Scholz R.-D., Bihain G., Storm J., 2014, Astronomy \& Astrophysics, 567, A43

\bibitem[{Schulte {et~al}\mbox{.}(2010)Schulte, Alegret, Arenillas, Arz,
  Barton, Bown, Bralower, Christeson, Claeys, Cockell, {et~al.}}]{schulte10}
Schulte P. {et~al.}, 2010, Science, 327, 1214

\bibitem[{{Th{\'e}venin} {et~al}\mbox{.}(2002){Th{\'e}venin}, {Provost},
  {Morel}, {Berthomieu}, {Bouchy}, \& {Carrier}}]{thevenin02}
{Th{\'e}venin} F., {Provost} J., {Morel} P., {Berthomieu} G., {Bouchy} F.,
  {Carrier} F., 2002, \aap, 392, L9

\bibitem[{Trujillo \& Sheppard(2014)}]{trujillo14}
Trujillo C.~A., Sheppard S.~S., 2014, Nature, 507, 471

\bibitem[{{Tsiganis} {et~al}\mbox{.}(2005){Tsiganis}, {Gomes}, {Morbidelli}, \&
  {Levison}}]{tsiganis05}
{Tsiganis} K., {Gomes} R., {Morbidelli} A., {Levison} H.~F., 2005, \nat, 435,
  459

\bibitem[{{van Leeuwen}(2007)}]{leeuwen07}
{van Leeuwen} F., 2007, \aap, 474, 653

\bibitem[{{Weissman}(1990)}]{weissman90nat}
{Weissman} P.~R., 1990, \nat, 344, 825

\bibitem[{{Weissman}(1996)}]{weissman96}
{Weissman} P.~R., 1996, in Astronomical Society of the Pacific Conference
  Series, Vol. 107, Completing the Inventory of the Solar System, {Rettig} T.,
  {Hahn} J.~M., eds., pp. 265--288

\bibitem[{{Weissman}(2007)}]{weissman07}
{Weissman} P.~R., 2007, in IAU Symposium, Vol. 236, IAU Symposium, {Valsecchi}
  G.~B., {Vokrouhlick{\'y}} D., {Milani} A., eds., pp. 441--450

\bibitem[{Wiegert \& Tremaine(1999)}]{wiegert99}
Wiegert P., Tremaine S., 1999, Icarus, 137, 84

\bibitem[{Wilhelms, McCauley \& Trask(1987)Wilhelms, McCauley, \&
  Trask}]{wilhelms87}
Wilhelms D.~E., McCauley J.~F., Trask N.~J., 1987, Washington: USGPO; Denver,
  CO (Federal Center, Box 25425, Denver 80225): For sale by the Books and
  Open-file Reports Section, US Geological Survey, 1987., 1

\end{thebibliography}
\end{document}